%% file: WCIS2019-answerIRF.tex
\documentclass[sigconf]{acmart}
\usepackage{graphicx}

\usepackage{booktabs} % For formal tables
\usepackage{multirow}
\usepackage{multicol}
 \usepackage[font=small,labelfont=bf]{caption} % make caption small
\usepackage{balance}
\usepackage{comment}
\usepackage{amsmath}
\usepackage{subcaption}
\usepackage{wrapfig}
\setcopyright{rightsretained}
\usepackage{etoolbox}
\usepackage{mathrsfs}
\usepackage{mathtools}

\acmDOI{10.475/123_4}

\acmISBN{123-4567-24-567/08/06}

\acmConference[]{WCIS'19}{July 2019}{Paris, France}
\acmYear{2019}
\copyrightyear{2019}

\acmPrice{15.00}

\begin{document}
	\title{Revisiting Iterative Relevance Feedback for Document and Passage Retrieval}

	\begin{comment}

		\author{~}
	
	\orcid{~}
	\affiliation{%
		\institution{~}
		\streetaddress{~}
		\city{~} 
		\state{~} 
		\postcode{~}
	}
	\affiliation{%
		\institution{~}
	}
	\email{ }

	\end{comment}	
	\author{Keping Bi}
	%\authornote{Dr.~Trovato insisted his name be first.}
	\affiliation{%
	\institution{University of Massachusetts}
	% \streetaddress{P.O. Box 1212}
	\city{Amherst} 
	\state{MA} 
	}
	\email{kbi@cs.umass.edu}
	
	\author{Qingyao Ai}
	\affiliation{%
	\institution{University of Massachusetts}
	\city{Amherst} 
	\state{MA} 
	}
	\email{aiqy@cs.umass.edu}
	
	\author{Bruce Croft}
	\affiliation{%
	\institution{University of Massachusetts}
	\city{Amherst} 
	\state{MA} 
	}
	\email{croft@cs.umass.edu}
	\begin{comment}
	\end{comment}

%\author{Keping Bi \inst{1} \and
%	Qingyao Ai \inst{1} \and
%	Bruce Croft \inst{1}}

%
\begin{abstract}
	As more and more search traffic comes from mobile phones, intelligent assistants, and smart-home devices, new challenges (e.g., limited presentation space) and opportunities come up in information retrieval. 
	Previously,  an effective technique, relevance feedback (RF), has rarely been used in real search scenarios due to the overhead of collecting users' relevance judgments.
	However, since users tend to interact more with the search results shown on the new interfaces, it becomes feasible to obtain users' assessments on a few results during each interaction. 
	This makes iterative relevance feedback (IRF) techniques look promising today. 
	IRF can deal with a simplified scenario of conversational search, where the system asks users to provide relevance feedback on results shown in the current iteration and shows more relevant results in the next interaction. 
	IRF has not been studied systematically in the new search scenarios and its effectiveness is mostly unknown. 
	In this paper, we re-visit IRF and extend it with RF models proposed in recent years.
	We conduct extensive experiments to analyze and compare IRF with the standard top-k RF framework on document and passage retrieval.
	Experimental results show that IRF is at least as effective as the standard top-k RF framework for documents and much more effective for passages. 
	This indicates that IRF for passage retrieval has huge potential and is a promising direction for conversational search based on relevance feedback. 

\end{abstract}

\begin{comment}
\begin{CCSXML}
	<ccs2012>
	<concept>
	<concept_id>10002951.10003317.10003338</concept_id>
	<concept_desc>Information systems~Retrieval models and ranking</concept_desc>
	<concept_significance>500</concept_significance>
	</concept>
	</ccs2012>
\end{CCSXML}

\ccsdesc[500]{Information systems~Retrieval models and ranking}
\end{comment}

\maketitle              % typeset the header of the contribution

\input{introduction}
\input{related-work}
\input{word-based-IRF}

\input{experiment-IRF}
\input{conclusion}

\bibliographystyle{ACM-Reference-Format}
\bibliography{WCIS2019-answerIRF} 

\end{document}

%% file: introduction.tex
%!TEX root=WCIS2019-answerIRF.tex
\section{Introduction}
\label{sec:introduction}

Recently, the interface of modern search engines has experienced significant changes.
More than 50\% of search traffic comes from mobile phones in 2018 \footnote{%Desktop vs Mobile vs Tablet Market Share Worldwide - September 2018 
\url{http://gs.statcounter.com/platform-market-share/desktop-mobile-tablet}}, and the number of people who use intelligent assistants (e.g., Siri) and smart-home devices (e.g., Echo) for search is also increasing today.
On the one hand, new search environments introduce new challenges to search engines. For example, the precision of top-1 results could significantly affect user experience because assistants or smart-home devices usually present only one result at a time.
On the other hand, the modern search scenarios provide new opportunities for the study of interactive search. 
People tend to interact more with phones and smart-home devices, so deploying relevance feedback (RF) techniques to real search systems becomes feasible and promising.

Relevance feedback has been shown to be effective through extensive studies in the IR community~\cite{salton90improvingretrieval,rocchio1971relevance,robertson1976relevance,zhai2001model,lavrenko2001relevance,brondwine2016utilizing}.
The idea of RF is to use the explicit relevance judgments provided by users to refine the query model and further retrieve more relevant results. 
Most existing studies focus on developing an effective RF model that improves the retrieval system in a single iteration, where users assess the relevance of top 10 or more documents in the initial ranking list~\cite{salton90improvingretrieval}. %, harman1992relevance
Due to significant manual efforts required for relevance judgments, these RF models have been seldom used in real search scenarios. 
%Such paradigm, however, may not be preferable in the new search environments.

\begin{figure}[h]
	\centering
	\includegraphics[width=0.8\linewidth]{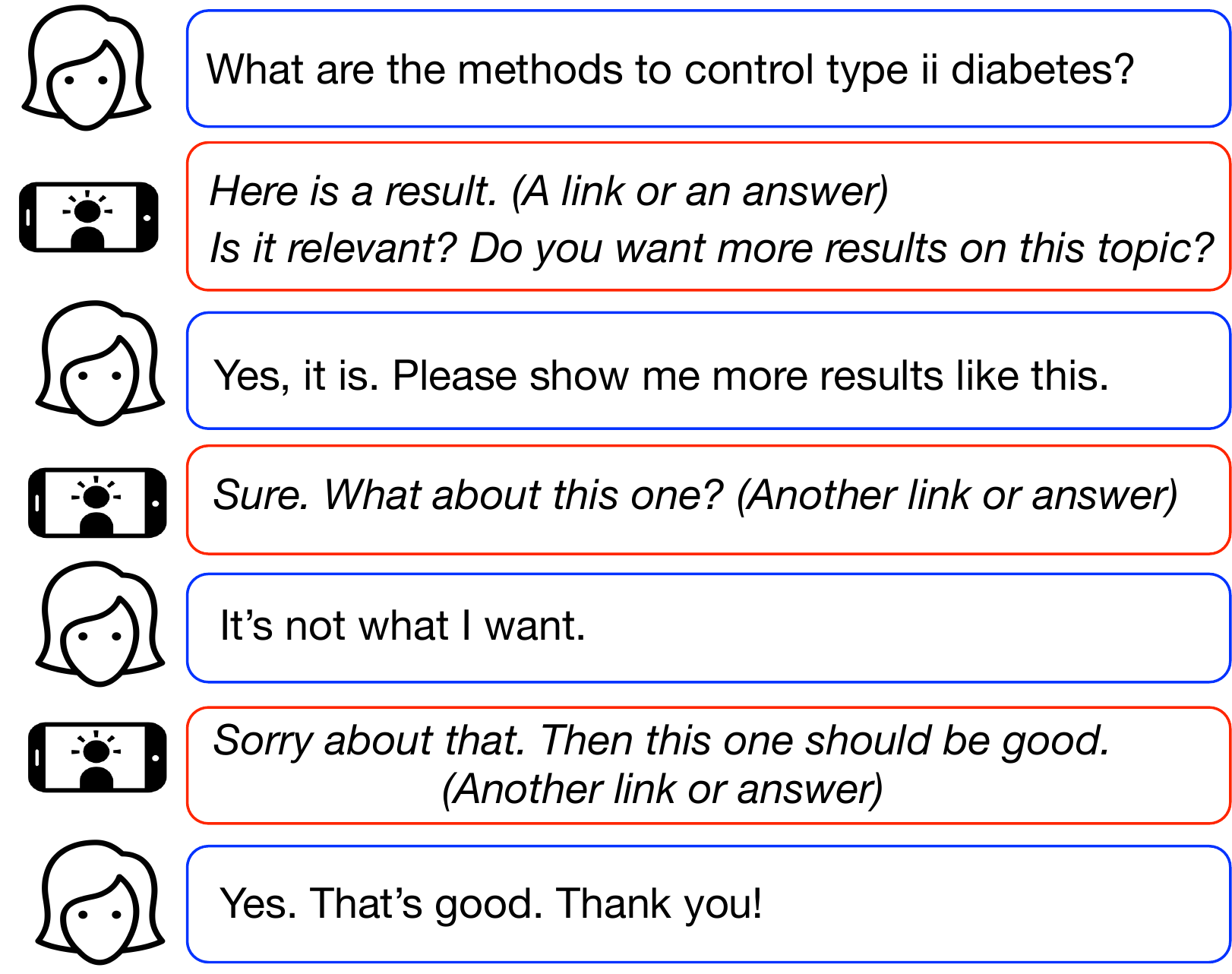}
	\caption{An example conversation on mobile devices where relevance feedback can be obtained and iterative search is preferred. }
	\label{fig:conv_irf}
\end{figure}

In new search environments, relevance feedback could be potentially collected through the interactions between users and the system. Figure \ref{fig:conv_irf} shows an example conversation between the assistant and a user where the quality of a search result can be obtained during the interaction. 
Since the display space or bandwidth is severely limited, it is more natural to do re-ranking iteratively after collecting user feedback on a small number of results during the search interactions rather than gathering a lot of feedback and do a one-shot retrieval refinement.
We refer to the former as iterative relevance feedback (IRF) and the latter as the standard top-k RF framework.

After IRF was proposed and investigated based on Rocchio in the 1990s, many new RF techniques~\cite{lavrenko2001relevance,zhai2001model,brondwine2016utilizing} have appeared. 
However, as far as we know, there is no systematical study on IRF techniques 
%Despite its potential for new search interfaces, as far as we know, there is little research on IRF since the 1990s. In the twenty years after IRF was proposed, many new RF techniques~\cite{lavrenko2001relevance,zhai2001model} have appeared. Moreover, mobile and voice search has become a trend. 
and the effectiveness of IRF with new RF methods in modern search scenarios remains mostly unknown.
In this paper, we conduct a systematic study of IRF with more recent models and under different scenarios.
Specifically, we focus on two research questions: 
(1) Given a fixed budget (relevance judgments), does IRF perform better than the standard top-k RF framework for recent RF methods?
(2) Does IRF perform equally well in retrieval tasks with different granularities?
To answer these questions, we extend several representative RF methods to iterative versions and conduct extensive experiments on two search tasks, document and passage retrieval. 
The first task aims to simulate the cases where users conduct traditional ad-hoc retrieval (coarse granularity) with limited display space (e.g., phone screen), while the second task focuses on the scenarios where a search engine directly returns the answer or a relevant passage of a query (e.g., search on intelligent assistants or smart-home devices, fine granularity).
Experimental results show that IRF works better or at least similar to the standard top-k RF framework on document retrieval
and much more effective on passage collections. 
% In addition, it performs more effective on passage retrieval where results of finer granularity are retrieved.

%% file: related-work.tex
%!TEX root=WCIS2019-answerIRF.tex
\section{Related Work}
\label{sec:related_work}
\textbf{Relevance Feedback}. In general, there are three types of relevance feedback (RF) methods for ad-hoc retrieval, which are based on the vector space model (VSM)~\cite{salton1975vector}, the probabilistic model \cite{maron1960relevance} and the language model (LM) for information retrieval (IR) \cite{ponte1998language}. 
Rocchio ~\cite{rocchio1971relevance} is an RF model based on VSM, which refines the vector of a user query by bringing it closer to the center of relevant documents and further from the center of non-relevant documents. % in the vector space.
In the probabilistic RF method, expansion terms are scored according to the probability of their occurrence in relevant documents compared to non-relevant documents.
More recently, feedback techniques have been investigated extensively based on LM, among which, the relevance model \cite{lavrenko2001relevance} and the mixture model \cite{zhai2001model} are two well-known examples that empirically perform well \cite{lv2009comparative}.
Later, the Distillation model \cite{brondwine2016utilizing} adds a query specific non-relevant language models to the mixture model. In addition, there have also been systematical studies on various pseudo RF methods in LM and VSM \cite{hui2011comparative,lv2009comparative}, but no such study on IRF.  

\textbf{Iterative Relevance Feedback}. IRF was first proposed by Aalsberg et al. \cite{aalbersberg1992incremental} based on Rocchio. 
In their work, users are asked to judge a single result shown in each interaction, then the query model can be refined iteratively with feedback. 
This approach showed better performance than standard batch feedback. 
Later, Allan et al. \cite{allan1996incremental} showed the effectiveness of incremental RF also based on Rocchio for information filtering.
Different from their work, we revisit IRF with recently proposed models on retrieval tasks of different granularities. 

Some recent TREC tracks \cite{yang2016trec,grossman2016trec} also made use of IRF, but their objectives are different and require a large amount of user feedback.
The Total Recall track  \cite{yang2016trec} aims to promote all of the relevant documents before non-relevant ones with a huge number of relevance judgments. 
The target of the Dynamic Domain track \cite{grossman2016trec} is to identify documents satisfying all the aspects of the users' information need with passage-level feedback. 
In contrast,  we investigate IRF with a fixed small amount of feedback and perform a systematical study of IRF for both document and passage retrieval. 
%As far as we know, no study has systematically investigated the performance of IRF with recently proposed retrieval models for both document and passage retrieval.

%TODO: TREC dynamic domain track and total recall track.
%TODO: still emphasize no systematic study was done on different granularities of data.

%% file: word-based-IRF.tex
%!TEX root=WCIS2019-answerIRF.tex
\section{Iterative RF Models}
\label{sec:iter_model}
In contrast to top-k RF, in iterative RF, on the one hand, better results may be identified within fewer iterations due to earlier re-ranking, which will reduce the cost of user assessment during search interactions. On the other hand, there are fewer results available with feedback, especially in the first several iterations. RF models require sufficient text to estimate the probabilities or weights of expansion terms that represent the relevance topic model accurately. Little text may be insufficient to distill the non-relevant topics contained in the relevant results and cause topic drift. This problem could be more severe for passages since they are shorter than documents. 

To study top-k RF and IRF systematically, we first reformulate some representative RF models based on the language model (LM), the vector space model (VSM) and the probabilistic framework as iterative models.
Specifically, we use the relevance model (RM3) \cite{lavrenko2001relevance} and the Distillation model \cite{brondwine2016utilizing} for LM~\cite{ponte1998language}; Rocchio \cite{rocchio1971relevance} for VSM~\cite{salton1975vector}; and a conventional method with adjusted deviation (Prob) for the probabilistic framework~\cite{salton90improvingretrieval}.
To generate the initial ranking, we use Query Likelihood (QL) for LM, BM25 for VSM and Prob.

%In the application scenario of IRF, we collect user feedback on a fixed number of top retrieved results in each iteration.
To keep the query model from diverging to non-relevant topics, we maintain two pools for relevant and non-relevant results.
Let $RP^{(i)}$ and $NRP^{(i)}$ be the set of all the judgments collected until the $i$th iteration.
Then, in the $i$th iteration, new judged relevant results $R^{(i)}$ and non-relevant results $NR^{(i)}$ are added to $RP^{(i)}$ and $NRP^{(i)}$, i.e., 
$$RP^{(i)} = RP^{(i-1)} \cup R^{(i)}, ~~~~ NRP^{(i)} = NRP^{(i-1)} \cup NR^{(i)} $$
where $i > 0, RP^{(0)} = \emptyset, NRP^{(0)} = \emptyset$.
We also tried to incrementally estimate the query model in the $i$th iteration, i.e. $Q^{(i)}$, with $Q^{(i-1)}$, $R^{(i-1)}$ and $NR^{(i-1)}$.
This method, however, suffers from topic drift severely and performs much worse than using the original query $Q^{(0)}$ and $RP^{(i-1)}$ and $NRP^{(i-1)}$. 

\textbf{Iterative Relevance Model.}
RM3~\cite{lavrenko2001relevance} is a well-known pseudo RF method that has also been used for RF. 
Let $c(w, x)$ be the count of term $w$ in a piece of text $x$, and $p_x^{MLE}(w)=\frac{c(w, x)}{\sum_{w'\in x}{c(w',x)}}$ be the maximum likelihood estimate (MLE) of $w$ with respect to $x$.
% When the true relevant judgments are available, we can directly estimate the relevance model $p_{rel_{rm3}}(\cdot)$ with binary weights \cite[p. 69]{Lavrenko:2004:GTR:1087151} in the $i$th iteration ($i > 1$) according to
The relevance model in the $i$th iteration ($i > 1$)can be estimated with the true RF version of RM3 \cite[p. 69]{Lavrenko:2004:GTR:1087151} according to
\begin{equation}
p_{rel_{rm3}}^{(i)}(w) = \frac{1}{|RP^{(i - 1)}|} \sum_{x\in RP^{(i - 1)}} p_{x} ^{MLE}(w)
\end{equation}
Then, the updated query language model in the $i$th iteration is the linear combination of the original query language model $p_{Q^{(0)}}^{MLE}(\cdot)$ and $p_{rel_{rm3}}^{(i)}(\cdot)$.
\begin{comment}
i.e.,
\begin{equation}
\label{eq:rm3_q}
\resizebox{.7 \linewidth}{!} 
{
$p_{rm3, Q^{(i)}}(w)\stackrel{def}{=} \lambda_{rm3} p_{Q^{(0)}}^{MLE}(w) + (1-\lambda_{rm3}) p_{rel_{rm3}}^{(i)}(w)$
}
\end{equation}
%where $\lambda_{rm3}$ is a hyper-parameter ranging from 0 to 1.  
\end{comment}
Finally, the documents are ranked with the KL divergence between the  language models of the query and the documents. 
\begin{comment}
 which is   
\begin{equation}
\resizebox{.7 \textwidth}{!} 
{
$\begin{split}
score_{KL}(p_{rm3, Q^{(i)}}(\cdot), p_x^{Dir}(\cdot)) & \!\!=\!\! -KL(p_{rm3, Q^{(i)}}(\cdot) || p_x^{Dir}(\cdot)) \\
& \!\!\propto\!\! \sum_w{p_{rm3, Q^{(i)}}(w)  \log p_x^{Dir}(w)} 
\end{split}
$
}
\end{equation}
where $p_x^{Dir}(w)$ is the probability of $w$ from a Dirichlet smoothed unigram language model induced from a document $x$.
\end{comment}
%$score_{KL}(p_{rm3, Q^{(i)}}(\cdot), p_x^{Dir}(\cdot))$ 

\textbf{Iterative Distillation Model. } 
Distillation~\cite{brondwine2016utilizing} is one of the most recent RF methods, which extends the mixture model \cite{zhai2001model} by incorporating a query specific non-relevant topic model.
It assumes that terms in relevant documents are generated from a mixture of a relevance topic model $p_{rel_{distill}}(\cdot)$, a query specific non-relevance topic model $p_{NR}^{MLE}(\cdot)$, and a background corpus language model $p_C^{MLE}(\cdot)$. %The initial results given the original query $Q^{(0)}$ are ranked with QL according to $score_{KL}(p_{Q^{(0)}}^{MLE}(\cdot), p_{x}^{Dir}(\cdot))$. 
For the $i$th iteration ($i > 1$), $p_{rel_{distill}}^{(i)}(\cdot)$ is estimated with the EM algorithm to maximize the log likelihood of words in $RP^{(i-1)}$, i.e.,
\begin{comment}
\begin{equation}
\label{eq:distill}
\resizebox{.93 \textwidth}{!} 
{$
\sum_{x\in RP^{(i-1)}}\sum_{w}c(w,x) \log  \big( (1-\lambda_1 -\lambda_2) p_{rel_{distill}}^{(i)}(w) 
+ \lambda_1 p_{NRP^{(i-1)}}^{MLE}(w) + \lambda_2 p_C^{MLE}(w) \big) 
$}
\end{equation}
\end{comment}
\begin{equation}
\label{eq:distill}
\begin{split}
	\sum_{x\in RP^{(i-1)}}\sum_{w}c(w,x) \log & \big( (1-\lambda_1 -\lambda_2) p_{rel_{distill}}^{(i)}(w) \\
	+ & \lambda_1 p_{NRP^{(i-1)}}^{MLE}(w) + \lambda_2 p_C^{MLE}(w) \big) 
\end{split}
\end{equation}
where $\lambda_1$ and $\lambda_2$ are hyper-parameters. 
Note that if $\lambda_1$ is set to 0, Distillation is exactly the same as the mixture model~\cite{zhai2001model}. 
Similar to RM3, $p_{rel_{distill}}^{(i)}(\cdot)$ is linear combined with $p_{Q^{(0)}}^{MLE}(\cdot)$ to calculate the new query model for the $i$th iteration, which then acts as a basis to score results according to KL divergence. 
\begin{comment}
Similar to RM3, the new query model for the $i$th iteration is computed as: 
\begin{equation}
\label{eq:distill_q}
p_{distill, Q^{(i)}}(w)\!\stackrel{def}{=}\! \lambda_{distill} p_{Q^{(0)}}^{MLE}(w)\! +\! (1\!-\!\lambda_{distill}) p_{rel_{distill}}^{(i)}\!\!\!\!\!(w)
\end{equation}
The final results are ranked with $score_{KL}(p_{distill, Q^{(i)}}(\cdot), p_x^{Dir}(\cdot))$ 
%change x to X and state X refers to a document or a passage
\end{comment}

\textbf{Iterative Rocchio Model}.
In VSM, queries and documents are represented with vectors in high-dimensional term space. The weight of each dimension can be calculated in many ways and a similarity measure is used to score documents. In this work, we use the BM25 \cite{robertson1995okapi} weight for terms in a document or passage vector $\vec{x}$ and dot product as the similarity measure.
\begin{comment}
The BM25 weight is 
\begin{equation}
\frac{(k_1+1) \cdot c(w,x)}{k_1(1-b + b \frac{|x|}{avgdl})+c(w, x)} \cdot log \frac{|C|+1}{df_C(w)}
\end{equation}
where $C$ is the corpus; $|x|$ is the length of $x$; $avgdl$ is the average document length in $C$; $df_S(w)$ is the document frequency of $w$ in the set $S$ (which is $C$ in this case); and $k_1$, $b$ are hyper-parameters. 
\end{comment}
The term weight in the vector of the initial query $Q^{(0)}$ is set to be the term count in $Q^{(0)}$, i.e., $c(w,Q)$. 
%Initial results in the first iteration are retrieved using BM25 \cite{robertson1995okapi}. 
Then, the query vector in the $i$th iteration ($i>0$) is computed as 
\begin{equation}
\label{eq:rocchio}
\vec{Q^{(i)}} = \vec{Q^{(0)}} + \beta \frac{1}{|RP^{(i - 1)}|} \!\!\sum_{x\in RP^{(i - 1)}}\!\!\!\!\!\!\!{\vec{x}} + \gamma \frac{1}{|NRP^{(i - 1)}|} \!\!\!\sum_{x\in NRP^{(i - 1)}}\!\!\!\!\!\!\!\!\!{\vec{x}}
\end{equation}
\begin{comment}
\begin{equation}
\label{eq:rocchio}
\resizebox{.93 \textwidth}{!} 
{$
\vec{Q^{(i)}} = \vec{Q^{(0)}} + \beta \frac{1}{|RP^{(i - 1)}|} \sum_{x\in RP^{(i - 1)}}{\vec{x}} + \gamma \frac{1}{|NRP^{(i - 1)}|} \sum_{x\in NRP^{(i - 1)}}{\vec{x}}	
$}
\end{equation}
\end{comment}
where $\beta$ and $\gamma$ are the coefficients to balance the influence of postive and negative feedback.
%where $\vec{x}$ is the term frequency vector of a document $x$.
If $RP^{(i - 1)}$ or $NRP^{(i - 1)}$ is empty, the corresponding part is omitted. The relevance score of a document or an answer passage x with respect to a query is computed with the dot product between $\vec{Q^{(i)}}$ and $\vec{x}$.  
\begin{comment}
, namely,
\begin{equation}
score_{VSM}(Q^{(i)}, x) = \vec{Q^{(i)}} \cdot \vec{x} 
\end{equation}
\end{comment}

\textbf{Iterative Probabilistic Model}.
%In the probabilistic model, expansion terms are selected according to the probability they are from relevant documents compared to non-relevant documents \cite{robertson1976relevance}. %, harman1992relevance}. 
In the probabilistic framework \cite{robertson1976relevance}, the feedback model at $i$th iteration is estimated by
\begin{equation}
\label{eq:prob}
\begin{split}
& p_{prob}^{(i)}(w) = \log \big (p_w(1-u_w) / u_w(1-p_w) \big) \\
& p_w = P(w|rel) = \frac{df_{RP^{(i-1)}}(w)+df_C(w) / |C|}{|RP^{(i-1)}| + 1} \\
& u_w = P(w|nonrel) = \frac{df_C(w) - df_{RP^{(i-1)}}(w) + df_C(w) / |C|}{|C| - |RP^{(i-1)}| + 1} \\
\end{split}
\end{equation}
where $df_S(w)$ is the document frequency of $w$ in the set $S$ (corpus $C$ and $RP^{(i-1)}$ in this case); 
The term weight of the original query is computed with
\begin{equation}
p_{prob, Q^{(0)}}(w) = \log \big ( (|C| - df_C(w))/df_C(w) \big)
\end{equation}
The query model at $i$th iteration is the linear combination between $p_{prob,Q^{(0)}}(\cdot)$ and $p_{prob}^{(i)}(\cdot)$. Again, dot product is used to score documents or passages. 
\begin{comment}
\begin{equation}
\label{eq:prob_q}
p_{prob, Q^{(i)}}(w)\stackrel{def}{=} \lambda_{prob} p_{prob,Q^{(0)}}(w) + (1-\lambda_{prob}) p_{prob}^{(i)}(w)
\end{equation}
Then, the score of a result $x$ given the query model $p_{prob, Q^{(i)}}(w)$ is computed by the dot product of the query model and the document model where the weight of each term is estimated by $p_x^{MLE}(w)$. 
%$score_{prob}(Q^{(i)}, x) = \vec{Q^{(i)}} \cdot \vec{x} $. 
\end{comment}

%% file: experiment-IRF.tex
%!TEX root=WCIS2019-answerIRF.tex

\section{IRF Experiments}
\label{sec:irf_exp_doc_psg}
% In this section, we introduce the experimental setup and results of our iterative feedback models on document and answer passage retrieval. 
\begin{table} [t]
	\centering
	\small
	\caption{Statistics of experimental datasets.}
	\label{tab:dataset}
	%    \addtolength{\tabcolsep}{3pt} 
	%\scalebox{0.9}{    
	\begin{tabular}{ l || l |  l  | l | l | l } %p{5mm}
		\hline
		Dataset & \#Docs & DocLen & Vocab & \#Query & \#Qrels \\ \hline
		Robust & 0.5M & 504 & 0.6M & 250 & 17,412 \\ 
		Gov2 & 25M & 893 & 35M & 150 & 26,917 \\  \hline
		WebAP & 379k & 45 & 59k & 80 & 3843 \\ 
		%    PsgRobust & 383k & 46 & 64k & 246 & 6589 \\ 
		\hline
	\end{tabular}
	%\vspace{-15pt}
	%}    
\end{table}

\subsection{Experimental Setup}\label{sec:IRF_metrics}
%\textbf{Experimental Setup.} \label{sec:IRF_metrics}
% \textbf{Data.}
We used standard TREC collections, Robust, Gov2, for document retrieval and WebAP \cite{Yang2016BeyondFQ,keikha2014retrieving} for passage retrieval. Statistics of the datasets are summarized in Table~\ref{tab:dataset}.
% Robust contains high-quality news articles and Gov2 consists of ``.gov'' domain web pages. 
% The titles of topics in Robust and Gov2 are used as queries.
% WebAP~\cite{Yang2016BeyondFQ,keikha2014retrieving} is built on Gov2 and has passage-level annotations on a subset of queries for non-factoid answer retrieval.
%TODO show dataset statistics
%\subsubsection{System Settings}\label{sec:irf_exp_setup}
% \textbf{System Settings}.
All the methods were implemented based on the Galago toolkit \footnote{http://www.lemurproject.org/galago.php}. 
Stopwords were removed and words were stemmed with Krovetz Stemmer. % using the standard INQUERY stopword list 
To compare IRF with typical top-k feedback in a fair manner, we fixed the total number of judged results as 10 and experimented on 1, 2, 5, and 10 iterations, where 10, 5, 2, 1 results were judged during each iteration, respectively. 
Then, $10D \times 1I$ (10Doc-1Iter) is exactly the top-k feedback. 
Considering the limitation of presenting results in a real interactive search scenario, we pay more attention to the settings of one or two results per iteration.
Users' judgments were simulated by true labels of results.

All the parameters were set using 5-fold cross-validation with grid search. %except for retrieval with QL and BM25 on Robust and Gov2, where we use the average value of 5-cross validation from Huston et al's experiments \cite{huston2014comparison}. 
%For WebAP and PsgRobust, 
We tuned $\mu$ of QL in $\{30,50,300,500,1000,1500\}$ and $k$ of BM25 from $\{1.2,1.4,\cdots,2\}$. b is set as 0.75.
% as suggested by \cite{mogotsi2010christopher}. 
% where $\mu$ is 934 and 1481 repectively for QL
We scanned $\lambda_1$, $\lambda_2$ in Equation \ref{eq:distill} and the interpolation coefficient for the feedback model from $\{0, 0.2, 0.4, \cdots, 1.0 \}$,
%$\lambda_{rm3}$, $\lambda_{distill}$ and $\lambda_{prob}$ in Equation \ref{eq:rm3_q}, \ref{eq:distill_q} and \ref{eq:prob_q} 
%(the sum of $\lambda_1$ $\lambda_2$ should be less than 1), 
%$\lambda_{rm3}$ and $\lambda_{distill}$ in Equation \ref{eq:rm3_q} and \ref{eq:distill_q}  from $\{0, 0.2, 0.4, \cdots, 0.8 \}$, 
the number of expansion terms $m$ from $\{10, 20, \cdots, 50 \}$, and $\beta$, $\gamma$ in equation \ref{eq:rocchio} from $\{0, 0.5, 1, \cdots, 3.0 \}$. 

% \textbf{Evaluation.}
%The evaluation should only focus on the ranking of unseen results. So 
Similar to \cite{aalbersberg1992incremental}, we use freezing ranking \cite{cirillo1969evaluation} to evaluate the performance of IRF.
%ruthven2003survey
%and standard (regular) ranking list, as in \cite{brondwine2016utilizing,}.  
%RF focuses on improving the ranking of results which have not been examined so far. The performance of RF is usually evaluated in terms of the residual collection method, or the freezing ranking, \cite{cirillo1969evaluation,ruthven2003survey}. 
%The \textsl{freezing ranking} paradigm freezes the ranks of all results presented to the user in the earlier feedback iterations and assigns the first result retrieved in the $i$th iteration rank $iN+1$, where $N$ is the number of results shown in each iteration. 
%Note that previously shown results are filtered out in the following retrieval and results retrieved in the last iteration are appended to make a longer rank list.
The result lists are formed according to the order they are shown to users during the interactions. Previously shown results are removed in the following retrieval. Results retrieved in the last iteration are appended to the final rank list.
Then we use $MAP$ at cutoff 1000 and $NDCG@20$ to measure the performance of results overall and on the top. As suggested by Smucker et al. \cite{smucker2007comparison}, Fisher randomization test with threshold 0.05 is used to calculate statistical significance.

\begin{table*}
	\caption{Performance of iterative feedback on document and answer passage collections. * and + denote significant improvements over the initial ranked list (Initial) and the standard top-10 feedback model ($10\times1$). The initial ranking model is QL for RM3, Distillation, and BM25 for Rocchio and Prob. Best $MAP$ and $NDCG$ of each method are marked in bold.}
	\label{tab:word_irf_doc_psg}
	\small
	%    \addtolength{\tabcolsep}{3pt} 
	\scalebox{1}{    
	
	\begin{tabular}{  p{1.2cm} | c || l || l ||  l || l | l | l || l || l ||  l | l | l  }
		%\begin{tabular}{ c || c | c | c | c  }
		\hline
		\multicolumn{2}{c||}{\multirow{3}{*}{Dataset}} & Method & \multicolumn{5}{c||}{$MAP$ of freezing rank lists} & \multicolumn{5}{c}{$NDCG@20$ of freezing rank lists}\\
		\cline{4-13}
		\multicolumn{2}{c||}{}& (Doc$\times$Iter) & Initial & (10$\times$1) & (5$\times$2) & (2$\times$5) & (1$\times$10) & Initial & (10$\times$1) & (5$\times$2) & (2$\times$5) & (1$\times$10) \\
		\hline
		%\cmidrule{3-6}
		\multirow{8}{0pt}{Document Retrieval} &\multirow{4}{*}{Robust}
		& RM3 & 0.253 & 0.316$^{*}$ & 0.321$^{*+}$ & 0.321$^{*+}$ & \textbf{0.324$^{*+}$} & 0.416 & 0.461$^{*}$ & 0.474$^{*+}$ & \textbf{0.478$^{*+}$} & \textbf{0.478$^{*+}$} \\ \cline{3-13}
		&& Distillation & 0.253 & 0.311$^{*}$ & 0.321$^{*+}$ & 0.322$^{*+}$ & \textbf{0.327}$^{*+}$ & 0.416 & 0.461$^{*}$ & 0.474$^{*+}$ & 0.480$^{*+}$ & \textbf{0.486}$^{*+}$ \\ \cline{3-13}
		&& Rocchio & 0.255 & 0.316$^{*}$ & \textbf{0.325}$^{*+}$ & 0.315$^{*}$ & 0.316$^{*}$ & 0.418 & 0.463$^{*}$ & \textbf{0.476}$^{*+}$ & 0.462$^{*}$ & 0.467$^{*}$ \\ \cline{3-13}
		&& Prob & 0.255 & 0.287$^{*}$ & 0.287$^{*}$ & \textbf{0.288*} & 0.287$^{*}$ & 0.418 & 0.451$^{*}$ & 0.450$^{*}$ & \textbf{0.456}$^{*+}$ & 0.455$^{*}$\\
		\cline{2-13}
		&\multirow{4}{*}{Gov2}
		& RM3 & 0.294 & \textbf{0.349}$^{*}$ & 0.343$^{*}$ & 0.338$^{*}$ & 0.337$^{*}$ & 0.405 & 0.451$^{*}$ & \textbf{0.464}$^{*+}$ & 0.454$^{*}$ & 0.458$^{*+}$\\ \cline{3-13}
		&& Distillation & 0.294 & \textbf{0.339}$^{*}$ & 0.337$^{*}$ & 0.336$^{*}$ & \textbf{0.339}$^{*}$ & 0.405 & 0.443$^{*}$ & 0.452$^{*+}$ & 0.452$^{*+}$ & \textbf{0.464}$^{*+}$ \\ \cline{3-13}
		&& Rocchio & 0.295 & 0.316$^{*}$ & \textbf{0.327}$^{*+}$ & 0.323$^{*+}$ & 0.326$^{*+}$ & 0.416 & 0.447$^{*}$ & \textbf{0.456}$^{*+}$ & 0.453$^{*+}$ & 0.450$^{*}$ \\ \cline{3-13}
		&& Prob & 0.295 & \textbf{0.317}$^{*}$ & 0.316$^{*}$ & 0.314$^{*}$ & 0.315$^{*}$ & 0.416 & 0.442$^{*}$ & \textbf{0.454}$^{*+}$ & 0.450$^{*+}$ & 0.450$^{*+}$ \\
		\hline
		\hline
		\multirow{4}{0pt}{Passage Retrieval} &\multirow{4}{*}{WebAP}
		& RM3  & 0.093 &  0.115$^{*}$  & 0.121$^{*+}$  & \textbf{0.132$^{*+}$}  & 0.130$^{*+}$  & 0.143 & 0.166$^{*}$ & 0.174$^{*+}$ & 0.186$^{*+}$ & \textbf{0.189}$^{*+}$ \\ \cline{3-13}
		&& Distillation & 0.093 & 0.118$^{*}$ & 0.115$^{*}$ & \textbf{0.134}$^{*+}$ & 0.132$^{*+}$ & 0.143 & 0.166$^{*}$ & 0.177$^{*+}$ & 0.185$^{*+}$ & \textbf{0.187}$^{*+}$ \\ \cline{3-13}
		&& Rocchio & 0.101 & 0.120$^{*}$ & 0.134$^{*+}$ & 0.138$^{*+}$ & \textbf{0.139}$^{*+}$ & 0.150 & 0.167$^{*}$ & 0.179$^{*+}$ & 0.183$^{*+}$ & \textbf{0.185}$^{*+}$ \\ \cline{3-13}
		&& Prob & 0.101 & 0.127$^{*}$ & 0.130$^{*}$ & 0.134* & \textbf{0.138}* & 0.150 & 0.170$^{*}$ & 0.177$^{*+}$ & 0.183$^{*+}$ & \textbf{0.192}$^{*+}$ \\
		\hline            
	\end{tabular}  
	}
	%    \vspace{-10pt}
\end{table*}

\begin{comment}
\cline{2-13}
&\multirow{4}{*}{PsgRobust}
& RM3  & 0.280 & 0.322$^{*}$  & 0.328$^{*+}$  & 0.334$^{*+}$  & \textbf{0.335}$^{*+}$  & 0.319 & 0.356$^{*}$ & 0.362$^{*+}$ & 0.370$^{*+}$ & \textbf{0.372}$^{*+}$ \\  \cline{3-13}
&& Distillation & 0.280 & 0.319$^{*}$ & 0.326$^{*+}$ & 0.337$^{*+}$ & \textbf{0.340}$^{*+}$ & 0.319 & 0.354$^{*}$ & 0.360$^{*+}$ & 0.372$^{*+}$ & \textbf{0.376}$^{*+}$\\ \cline{3-13}
&& Rocchio & 0.219 & 0.296$^{*}$ & 0.310$^{*+}$ & 0.315$^{*+}$ & \textbf{0.319}$^{*+}$ & 0.292 & 0.342$^{*}$ & 0.358$^{*+}$ & 0.362$^{*+}$ & \textbf{0.366}$^{*+}$ \\ \cline{3-13}
&& Prob & 0.219 & 0.279$^{*}$ & 0.288$^{*+}$ & 0.295$^{*}$ & \textbf{0.296}* & 0.292 & 0.332$^{*}$ & 0.348$^{*+}$ & 0.352$^{*+}$ & \textbf{0.353}$^{*+}$\\
\end{comment}

\subsection{Results and Discussion}
% \label{sec:exp_result_IRF_doc_answer}
%\textbf{Results and Discussion.}
In this section, we discuss and compare the performance of IRF and standard top-k RF feedback in retrieval tasks with different granularities.
\begin{comment}
In contrast to standard top-k RF, IRF performs re-ranking earlier after a small number of results are assessed in each iteration.  
On the one hand, better results may be identified with fewer iterations due to earlier re-ranking, which will reduce the cost of user assessment during search interactions.  
On the other hand, less feedback information in each iteration may be not enough to estimate an accurate query model and other topics mentioned in the relevant results may cause topic drift in the iterative process. 
%In this paper, we investigate IRF on both document and answer passage retrieval to analyze the above effects on tasks with different granularity.
\end{comment}
Table \ref{tab:word_irf_doc_psg} shows the performance of the initial rank lists (QL for RM3 and Distillation, BM25 for Rocchio and Prob), standard top-10 RF ($10\times1$) and the IRF experimental results ($5\times2$, $2\times5$, $1\times10$).
%All the values of feedback methods shown in Table \ref{tab:word_irf_doc_psg} are significantly better than their retrieval baselines, i.e. RM3 and Distillation compared with QL, Rocchio and Prob compared with BM25, both on $MAP$ and on $NDCG@20$.
In general, IRF is effective on both document and passage collections in most cases.

%A
%document retrieval no clear clue of the performance with the number of iterations.
%not for sure to be better
%the reason.
%B
%Robust better than Gov2
%possible reason 
%C
%top result still better, maybe 
For document retrieval, IRF improves the performance compared with the top-k framework under many iteration settings, but there is no clear correlation between the performance with the number of iterations. 
This indicates that increasing iteration numbers with a small amount of feedback in each iteration does not always improve the performance. 
Because documents usually span multiple topics, reducing the number of feedback documents in each iteration makes the ranking system more vulnerable to drift to the non-relevant topics contained in the judged relevant documents. 
IRF needs enough relevant documents to estimate a robust query model for users' true information need in order to keep the topic from drifting.

The topic drift problem is more severe in Gov2 than in Robust. 
IRF improves $MAP$ significantly in many cases on Robust, but has similar or worse $MAP$ on Gov2.
The reason could be that Robust is a homogeneous dataset of high-quality news articles and shorter average document length, while Gov2 is a heterogeneous collection of noisy web pages and longer average document length. So more non-relevant information may appear in the judged relevant documents in Gov2 and topic drift is more likely to happen. 

Besides, IRF tends to have better performance on top results compared with the overall rank list on document collections. In more cases, $NDCG@20$ is improved by iterative models compared with $MAP$, especially on Gov2. This indicates that top-ranked results are less suffered from the topic drift problem than results with lower ranking scores.

In contrast to document retrieval, the benefits of IRF for passage retrieval is much more compelling.
In Table~\ref{tab:word_irf_doc_psg}, the performance of IRF on WebAP is positively correlated with the number of iterations.
Almost all methods achieve their best results with 10 iterations. 
Since answer passages are much shorter than documents, 
they are usually focused on a single topic and less likely to suffer from topic drift. 
As a result, the whole retrieval system can obtain more improvements when re-ranking is done in an earlier stage, even when we have a limited number of feedback passages.
This indicates that IRF techniques could have significant potential for answer passage retrieval.

%% file: conclusion.tex
%!TEX root=WCIS2019-answerIRF.tex

\section{Conclusion and Future Work}
\label{sec:conclusion}
We reformulate feedback models in the three main feedback frameworks as iterative models and investigated the performance of these IRF models on document and passage retrieval. 
Results show that IRF is at least as effective as standard top-k feedback for retrieving documents and is more powerful in finding answers.
%There is still space for IRF to improve the performance on passage retrieval since answer passages are short and may not have enough words to estimate the term weights in feedback models accurately.
For future work, we consider incorporating semantic information to complement word-based IRF models for passage retrieval. 
We also intend to study how to identify the first relevant answer in fewer iterations based on negative feedback. 
%Neural models will be investigated to further improve the performance.

%There is still space for IRF to improve the performance on passage retrieval, since answer passages may be too short to estimate accurate feedback models
%are short and may not have enough words to estimate the term weights in feedback models accurately.
%For future work, we consider incorporating semantic information to complement word-based IRF models for passage retrieval with neural models.  
%Neural models will be investigated to further improve the performance.

%Our method focuses more on user requests of ``more like this". We know diversity is also very important to provide users more informative results and we will take it into account in our future work. 
%In addition, we will consider IRF on answer passage retrieval with end-to-end neural models.
%We also intend to study how to do feedback with non-relevant passages alone or together with relevant passages to find more relevant answers in fewer iterations or use only the negative information to identify the first relevant answer in fewer iterations. 

\section{Acknowledgments}
This work was supported in part by the Center for Intelligent Information Retrieval and in part by NSF IIS-1715095. Any opinions, findings and conclusions or recommendations expressed in this material are those of the authors and do not necessarily reflect those of the sponsor.
\begin{comment}
\end{comment}